\documentclass[conference]{IEEEtran}
\IEEEoverridecommandlockouts

\usepackage{cite}
\usepackage{amsmath,amssymb,amsfonts}
\usepackage{algorithmic}
\usepackage{graphicx}
\usepackage{textcomp}
\usepackage{xcolor}
\usepackage{float} %
\usepackage{array}
\usepackage{caption}  
\usepackage{hyperref} 
\usepackage[numbers, square]{natbib}
\usepackage{xcolor}

\def\BibTeX{{\rm B\kern-.05em{\sc i\kern-.025em b}\kern-.08em
    T\kern-.1667em\lower.7ex\hbox{E}\kern-.125emX}}
\begin{document}

\title{Reinforcement Learning for Rate Maximization in IRS-aided OWC Networks\\
}

\author{\IEEEauthorblockN{Ahrar N. Hamad, Ahmad Adnan Qidan, Taisir E.H. El-Gorashi and Jaafar M. H. Elmirghani, }
\IEEEauthorblockA{\textit{ Department of Engineering, King’s College London, London, United Kingdom,} \\
\textit
\{{ahrar.hamad, ahmad.qidan, taisir.elgorashi, jaafar.elmirghani}\}@kcl.ac.uk}
}
\maketitle

\begin{abstract}
Optical wireless communication (OWC) is envisioned as one of the main enabling technologies of 6G networks, complementing radio frequency (RF) systems to provide high data rates. One of the crucial issues in indoor OWC  is service interruptions due to blockages that obstruct the line of sight (LoS)  between users and their access points (APs). Recently, reflecting surfaces referred to as intelligent reflecting surfaces (IRSs) have been considered to provide improved connectivity in OWC systems by reflecting AP signals toward users. In this study, we investigate the integration of IRSs into an indoor OWC system to improve the sum rate of the users and to ensure service continuity.  We formulate an optimization problem for sum rate maximization, where the allocation of both APs and mirror elements of IRSs to users is determined to enhance the aggregate data rate. Moreover, reinforcement learning (RL) algorithms, specifically Q-learning and SARSA algorithms, are proposed to provide real-time solutions with low complexity and without prior system knowledge. The results show that using RL algorithms achieves near-optimal solutions that are close to the solutions of mixed integer linear programming (MILP). The results also show that the proposed scheme achieves up to a 45\% increase in data rate compared to a traditional scheme that optimizes only the allocation of APs while the mirror elements are assigned to users based on the distance.
\end{abstract}


\section{Introduction}
The increasing growth of connected devices and the ongoing evolution of Internet-based applications have led researchers to investigate wireless communication technologies beyond traditional radio frequency (RF) systems. Optical wireless communication (OWC) has gained attention as a promising solution to address the limitations of RF spectrum shortage due to its large unregulated bandwidth, high security, and immunity to electromagnetic interference \cite{dang2020should}. Visible light communication (VLC) is a key technology of the OWC that uses light-emitting diodes (LEDs) for both illumination and data transmission.
However, 
 the performance of an indoor OWC system might deteriorate in scenarios where obstacles block the direct line of sight (LoS) between the transmitter and receiver. Therefore,  a large number of access points (APs) are needed to achieve full coverage where each optical AP illuminates a confined coverage area. \cite{hussein2015mobile}.

To improve the reliability of the wireless systems and mitigate the LoS blockage challenge, intelligent reflecting surfaces (IRSs) technology has been proposed to redirect the signals toward users when the LoS path is blocked, expanding the network connectivity \cite{9475160}. Compared to decode and forward (DF) and amplify-and-forward (AF) relays, IRSs do not generate any new signals, and therefore, they improve the performance of wireless networks in terms of spectral and energy efficiency compared to active relays \cite{di2020reconfigurable}. Furthermore, IRS offers high flexibility and superior compatibility for practical implementation as they can be easily attached and removed due to their low weight and conformal geometry \cite{zhou2020spectral}.

Recently, the integration of IRS has attracted interest in indoor OWC for its potential to expand the coverage, reduce the impact of LoS blockage, and enhance the achievable data rate. Integrating IRS into VLC has been explored through two primary methods: one utilizing mirror arrays and the other employing metasurfaces. A mirror array consists of an array of small passive and individually controllable mirrors, while a metasurface consists of sub-wavelength elements (meta-atoms) arranged in a planar array. In \cite{abdelhady2021visible}, the performances of these types of IRS  were studied in a VLC system. The results show that the effectiveness of power focusing is influenced by the number of reflecting elements and the dimensions of the source and reflector.
 In \cite{sun2021sum}, the implementation of a reconfigurable intelligent surface (RIS) was proposed in a VLC system, considering LoS and specular non-line-of-sight (NLoS) links. A discrete matrix transforming the problem into a binary programming formulation is introduced to simplify the resource allocation between the RIS and users only, and a low-complexity algorithm is developed to maximize the sum rate.  Furthermore, in \cite{Sun2022}, a resource allocation optimization problem was formulated to enhance the spectral efficiency in IRS-aided VLC system, where frozen variable and minorization-maximization algorithms were used to determine the  IRS coefficient, power allocation, and user association matrices. In \cite{Ahrar}, a resource allocation optimization problem was formulated to allocate APs and IRS mirrors to users using exhaustive search method.



In general, the optimization problems formulated in OWC systems have high complexity due to the need for a large number of APs to ensure coverage \cite{aboagye2021intelligent}. Therefore, practical algorithms are required to enable their applications in real-world deployment. Reinforcement learning (RL) algorithms have been adopted to address various challenges in wireless communication networks. RL algorithms are dynamic algorithms where an agent learns the optimal decision-making within an environment through iterative trial and error. RL has been considered to optimize resource allocation in multi-user OWC to maximize data rates, spectral and energy efficiency, or other performance metrics
\cite{elgamal2021q}, \cite{elgamal2021reinforcement}.

 In contrast to the work in \cite{Sun2022} and \cite{Ahrar},
this paper proposes a joint allocation of APs and IRS mirrors to users in an IRS-aided OWC network using RL. We formulate the joint optimization problem with the objective of maximizing the sum data rate. The joint optimization problem is also formulated as Markov decision process (MDP), and two RL algorithms Q-learning and SARSA, are applied to provide practical solutions. The system performance is evaluated under varying conditions of transmitted optical power and LoS blockage. The results demonstrate the optimality of the proposed RL algorithms and the ability of the proposed schemes to enhance the performance under different blockage scenarios.

The rest of this paper is structured as follows: In Section II, the system configuration is modeled, and in Section III, the optimization problem is formulated. Section IV explains the use of RL algorithms to solve the optimization problem. The results are given and discussed in Section V. Finally, conclusions and future work are presented in Section VI.

\begin{figure}[t]
    \centering
    \includegraphics[width=0.4\textwidth]{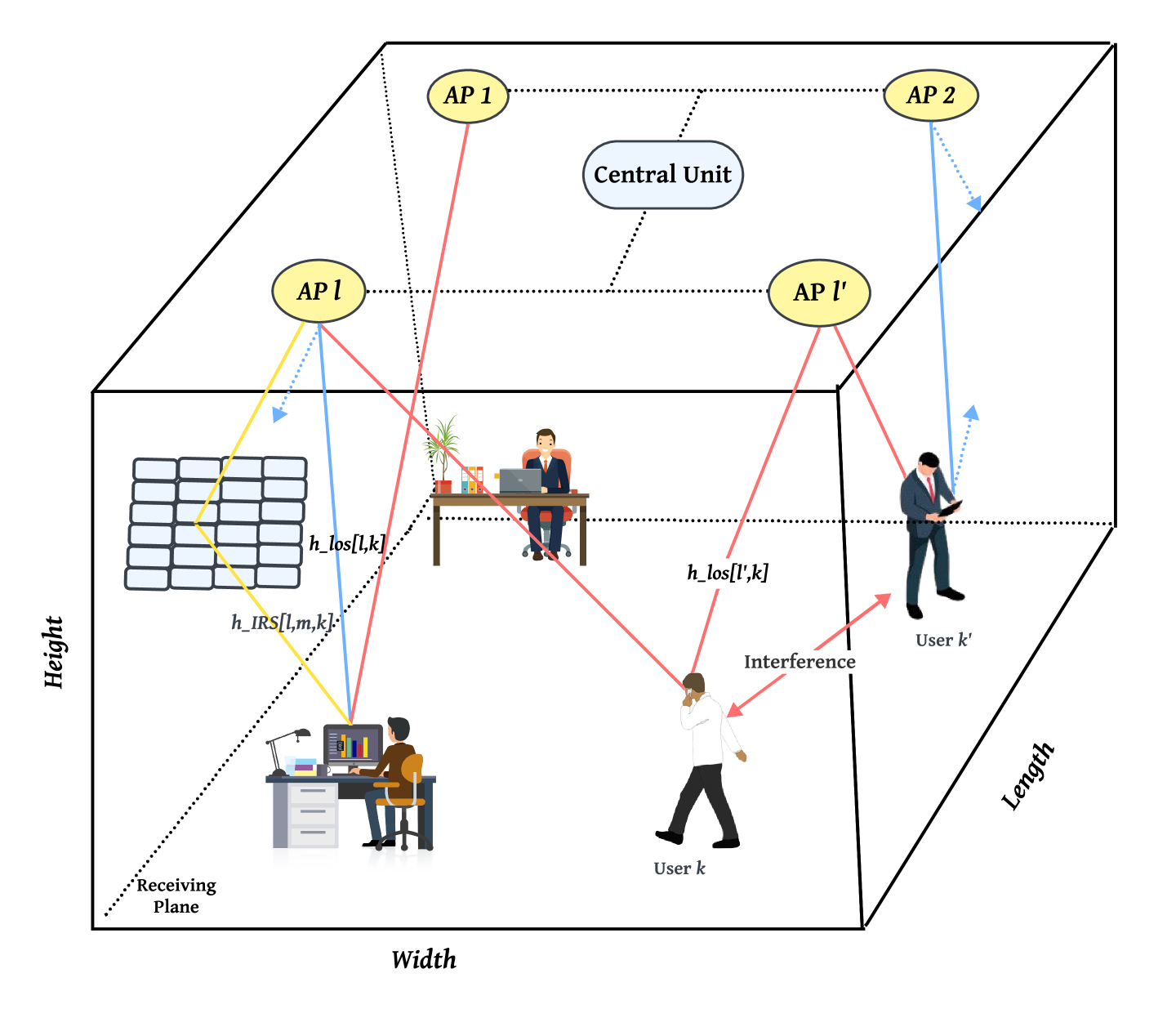}
     \captionsetup{justification=raggedright, singlelinecheck=false}
\caption{IRS-aided OWC system model.}
\label{fig}
\end{figure}
\section{SYSTEM MODEL} \label{sec:system}

 We consider a downlink VLC system where \( L \) optical APs are mounted on the ceiling to offer both illumination and communication for \( K \) users, who are randomly distributed across the communication plane, as illustrated in Fig. \ref{fig}. Each optical AP, $l \in L$, consists of multiple LEDs to provide a wider coverage area while maintaining eye safety regulations. Each user, $ k\in K$, is equipped with an optical receiver also known as angle diversity receiver (ADR) with multiple branch photodiodes which are oriented in different directions to provide independent channel responses. The directions of the ADR photodiodes are specified by their elevation \textit{($El$)} angles and azimuth\textit{($Az$)}. Further details on channel response calculations can be found in \cite{alsulami2020optimum}.
A mirror array, which plays the role of IRS with reflective elements, is mounted on the wall to improve the gain of the reflective channel. The mirror array consists of $M$, identical, passive, and smooth reflective mirrors. Each mirror,  $m \in M$, is rectangular with specific $width$ $\times$ $height$ dimensions and oriented according to two independent angles: the roll angle around the y-axis, $\epsilon_y$, and the yaw angle, $\vartheta_z$, around the z-axis.
Given the large number of mirrors in the indoor environment,  mirror orientations are determined to achieve better coverage in the room. Therefore, each user is expected to find at least one mirror available at its location that can improve its received signal.

Typically, VLC systems use intensity modulation/direct detection (IM/DD).To simplify the system, on-off keying (OOK) modulation is used. All the optical APs and IRSs as well as a WiFi AP installed for providing uplink transmission are connected to a central unit (CU) that has information on the network in terms of resource availability and users locations and demands. The CU uses this information to optimize the AP-user association and IRS mirror element allocation to maximize the sum rate of the network. 

In the following, the LoS link between a pair of an optical AP and a user, the specular NLoS link established through an IRS mirror, and the user data rate are derived in detail.

\subsection{The LoS channel gain}
The LoS component of the optical channel is the largest contributor to the power received by the user. The received channel gain by user \( k \in K\) from AP \( l \in L \) considering the LoS component is given as

\begin{equation}
h_{k,l}^{\text{LoS}} = 
\begin{cases} 
\frac{(n+1) A_r \cos^n(\alpha_{k,l}) \cos(\delta_{k,l})}{2 \pi D_{k,l}^2}, & 0 \leq \delta_{k,l} \leq \psi_c \\
\quad \quad 0, & \delta_{(k,l)} > \psi_c
\end{cases}
\end{equation}
where \( n \) is the order of Lambertian emission, which is based on the half-power semi-angle of the LED \(\phi_{1/2}\) and can be expressed as \( n = -\ln(2)/\ln(\cos(\phi_{1/2})) \). Moreover, \( A_r \) is the detector area, \( \alpha_{(k,l)} \) is the incident ray angle between the normal to the AP \( l \) and the irradiance ray of the user \( k \), \( \delta_{(k,l)} \) is the angle between the normal of the photodetector and the incident ray, and \( D_{(k,l)} \) is the distance between AP \( l \) and user \( k \). Note that, the incidence angle \( \delta_{(k,l)} \) must be within a range from 0 to the acceptance semi-angle of the concentrator (\( \psi_c \)), guaranteeing that the signal of the direct LoS is detected by the receiver; otherwise, no signal is received, i.e., $h_{k,l}^{\text{LoS}}$ = 0. 

\subsection{The IRS-reflected channel gain}

Assuming complete specular reflection by the mirrors of the IRS, the received channel gain by user \( k \in K\) from  mirror \( m \in M\) reflecting the signal from  AP \( l \in L \) is given as

\begin{equation}
h_{k,m,l}^{\text{IRS}} = 
\begin{cases} 
\frac{(n+1) \rho_m A_r dA_m \cos^n(\alpha_{m,l}) \cos(\beta_{k,m})}{2\pi (D_{m,l} + D_{k,m})^2}, & 0 \leq \beta_{k,m} \leq \psi_c \\
\quad \quad 0, & \beta_{k,m} > \psi_c
\end{cases}
\end{equation}
where \( \rho_m \) and \( dA_m \) are the reflection coefficient and the area of mirror \( m \), respectively, \( \alpha_{m,l} \) is the irradiance angle from AP \( l \) to mirror \( m \), \( \beta_{k,m} \) is the incidence angle of the signal reflected from mirror \( m \) to user \( k \), \( D_{m,l} \) is the distance between AP \( l \) and mirror \( m \) and \( D_{k,m} \) is the distance between mirror \( m \) and user \( k \). Moreover, the irradiance and incidence angles can be calculated as

\begin{equation}
\cos(\alpha_{m,l}) = \frac{\textbf{D}_{m,l} \cdot \textbf{n}_l}{\|\textbf{D}_{m,l}\|},
\end{equation}

\begin{equation}
\cos(\beta_{k,m}) = \frac{\textbf{D}_{k,m} \cdot \textbf{n}_k}{\|\textbf{D}_{k,m}\|},
\end{equation}
where \( \textbf{n}_l \) represents the normal vector to the AP plane, \( \textbf{n}_k \) denotes the normal vector at the receiver plane, \( \cdot \) is the inner product and \( \|\cdot\| \) denotes the Euclidean norm operators.
\subsection{User Data Rate}
In the system model considered, 
the signal received by user \( k \), $k \in K $, can be expressed as
\begin{equation}
y_k =  [ h_{k,l}^{\text{LoS}} + h_{k,m,l}^{\text{IRS}} ] P_k x_k + [ h_{k,l}^{\text{LoS}} + h_{k,m,l}^{\text{IRS}} ] P_{k'} x_{k'}+ z_k,
\label{resi}
\end{equation}

\noindent where \( x_t \) and \( P_t \), $t \in \{k,k'\}$, are the transmitted signal and power intended to a certain user. Note that multi-user interference can be avoided through serving users over exclusive frequencies or time slots. Moreover, \( z_k \) is the real-valued additive white Gaussian noise (AWGN) with zero mean and a certain variance. From equation \eqref{resi}, the signal-to-interference-plus-noise ratio $\textit{SINR}_k$ of user \( k \) can be expressed as
\begin{equation}
\textit{SINR}_k = \frac{[ R_0^2 P_k^2 (h_{k,l}^{\text{LoS}} + h_{k,m,l}^{\text{IRS}})^2 ]}{(I_k^2 + \sigma_t^2)}, 
\end{equation}

\noindent where \( R_0 \) is the receiver responsivity, \( I_k\) represents the interference received by user \( k \) from optical APs other than its corresponding optical AP, and \( \sigma_t^2 \) is the summation of preamplifier noise, shot noise associated with the received signal and background shot noise. Considering a tight lower bound of the achievable data rate for the dimmable VLC system derived in \cite{wang2013tight}, the data rate of user \( k \in K \) is given by

\begin{equation}
  R_k = \frac{B}{K_{in}} \log_2 \left( 1 + \frac{e}{2\pi} \textit{SINR}_k \right).
\end{equation}
where \( B \) is the modulation bandwidth, $ {K_{in} }$ denotes users interfering with each other, i.e., users served by the same optical APs,  and \( \frac{e}{2\pi} \) is for IM/DD.

\section{Problem Formulation} \label{sec:Pro}
In this section, an optimization problem is formulated to assign APs and IRS mirror elements to users in IRS-aided VLC systems to maximize the achievable sum rate. In this context, a logarithmic utility-based objective function is defined to maximize the sum data rate with proportional fairness among users. The objective function is given by

\[\text{(P1):} \quad
\max_{x,e} \sum_{k \in K} \log \left( \sum_{l \in L} \sum_{m \in M} x_{k,l}  e_{k,m}  R_k \right),
\]

\begin{subequations}
    \renewcommand{\theequation}{8\alph{equation}}
    \begin{equation}
        \text{s.t.}\quad R_k \geq R_{\text{min},k}, \quad \forall k \in K,
    \end{equation}
    \begin{equation}
        \sum_{l \in L} x_{k,l} = 1, \quad \forall k \in K,
    \end{equation}
    \begin{equation}
      \quad \quad x_{k,l} \in \{0,1\}, \quad \forall k \in K, \forall l \in L, 
    \end{equation}
    \begin{equation}
      \quad \sum_{m \in M} e_{k,m} = 1, \quad \forall k \in K,
    \end{equation}
    \begin{equation}
       \quad \quad e_{k,m} \in \{0,1\}, \quad \forall k \in K, \forall m \in M.
    \end{equation}\label{op}
\end{subequations}where $R_k$ is the data rate of user $k$,  \( x_{k,l} \) is a binary variable that equals 1 if AP \( l \) is allocated to user \( k \), otherwise, \( x_{k,l} = 0 \), and  \( e_{k,m} \) is a binary variable that sets the link from mirror \( m \) to user \( k \), i.e., if mirror \( m \) is allocated to user \( k \), \( e_{k,m} = 1 \), otherwise, \( e_{k,m} = 0 \). Moreover, constraint (8a) guarantees high QoS for each user,  and constraints (8b)  and (8d) ensure that each user receives useful information from a single pair of AP and mirror.

In \eqref{op}, as the number of APs, mirrors, and users increases, the number of possible allocations increases exponentially, making an exhaustive search for the optimal solution computationally intractable. Therefore,  we can divide the optimization problem into two sub-problems that can be solved separately. In the first sub-problem, APs are assigned to users based on LoS channel components as follows:

\begin{equation}
\label{OP2}
\begin{aligned}
\max_{x} \sum_{k \in K} \sum_{l \in L} \log \left( x_{k,l} R_{k}\right),\\
\textrm{s.t.} ~~~ (8a), (8b), (8c).~~~~~~~~\\
\end{aligned}
\end{equation}
The AP-user associations can be found through exhaustive search. In particular, the users report through the Wifi AP their locations and traffic demand to the CU. This information is then used to determine the AP-user associations considering the SNR and the transmission capacities of all the APs to satisfy the traffic demand. In the second sub-problem, mirrors are assigned to users based on IRS-reflected channel components considering the AP-user associations determined in the first sub-problem. In other words, the second sub-problem finds the mirror-user associations given the optimum assignment \( x_{k,l}^\dagger \) found by solving the first sub-problem. The second optimization problem can be rewritten as

\begin{equation}
\label{OP2}
\begin{aligned}
\max_{e} \sum_{k \in K} \sum_{l \in L} \sum_{m \in M} \log \left(x_{k,l}^\dagger e_{k,m} R_{k} \right),\\
\textrm{s.t.} ~~~ (8a), (8d), (8e).~~~~~~~~\\
\end{aligned}
\end{equation}
Note that, in real-time dynamic scenarios, solving the two sub-problems cannot provide solutions in real-time due to the high complexity of the optimization problems. In addition, full knowledge of the environment is required which might not be available. In the following, we propose RL algorithms to provide practical solutions.
\section{Intelligent Association using Reinforcement Learning} \label{sec:RL}
\begin{figure}[t]
    \centering
    \includegraphics[width=0.5\textwidth]{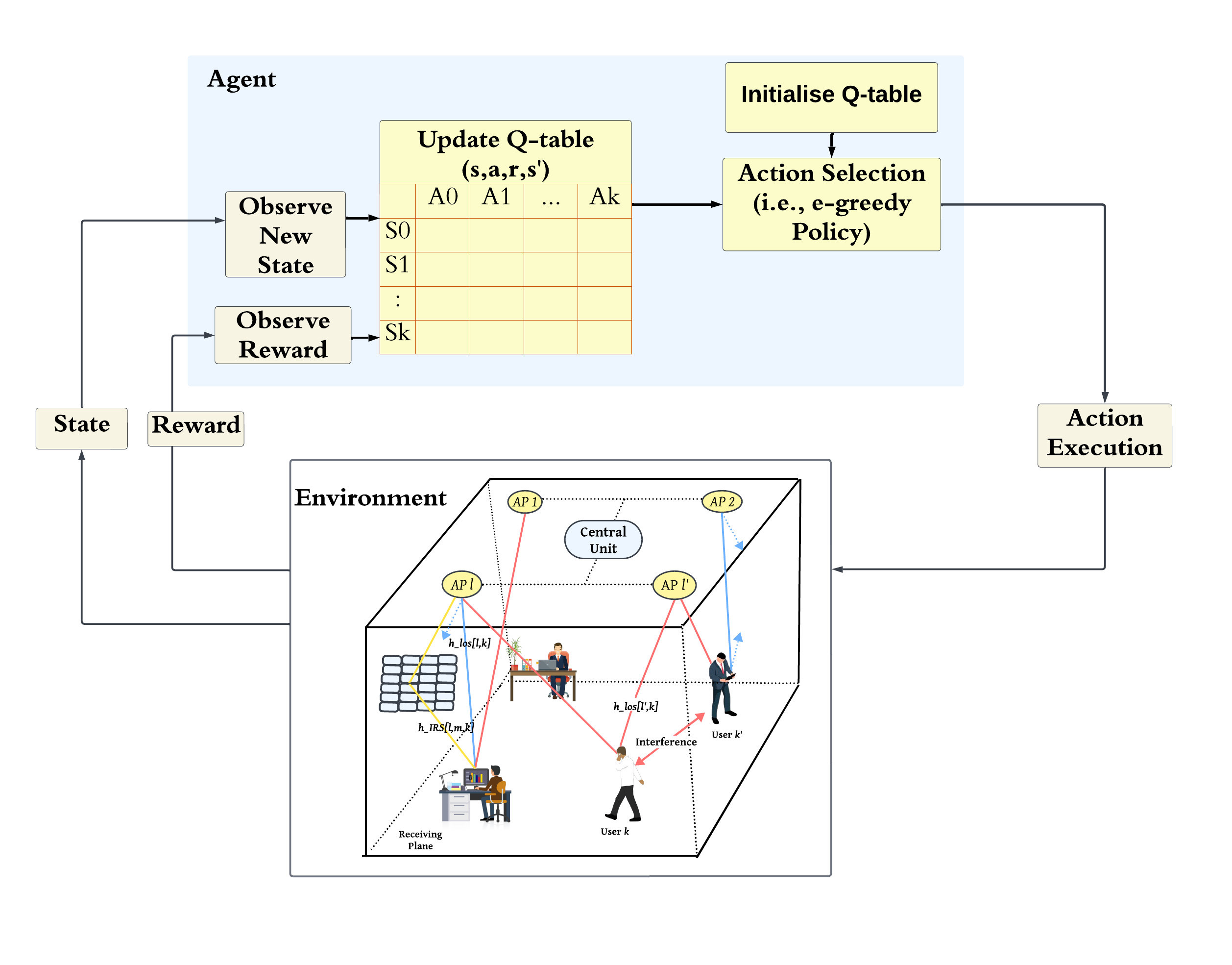}
\caption{Reinforcement learning for an indoor IRS-aided VLC system.}
\label{fig2}
\end{figure}

In RL, an agent learns to take actions by interacting with the network environment \cite{sutton2018reinforcement}. The first step in employing  RL is to frame the optimization problem as a MDP. This allows the RL agent to learn an optimal policy by interacting with its environment and receiving rewards. The MDP is a mathematical framework for modeling decision-making using five components: environment, agent, state space \( S \), action space \( \mathcal{A} \), and rewards \( r \). In our model, the environment is the IRS-VLC system and the CU acts as an agent making decisions about AP association and mirror allocation as shown in Fig \ref{fig2}.

The state space \( S \) is a finite set of discrete states, which represents the traffic demand of each user as in constraint  (8a), the SNRs for each user from all the APs and IRS mirrors, and the capacity of each AP. The action space \( \mathcal{A} \) is a finite set of discrete actions, which represents AP allocation and IRS mirrors allocation, i.e., \( x_{k,l} \) and \( e_{k,m} \). The CU controller selects an action from this space based on its observations from the environment and the actions must satisfy constraints  (8b) and (8e). Finally, the reward is the feedback that the agent receives after taking an action in a given state, and therefore, the reward function is designed to directly reflect high QoS and data rate as in \eqref{op}. That is 

\begin{equation}
r = \frac{x_{k,l} e_{k,m} R_k}{R_{\text{min},k}}
\end{equation}
Next, we employ two RL algorithms, Q-learning and SARSA, to solve the optimization problems in  \eqref{op}.

\subsection{Q-Learning} \label{sec:qlearning}

Q-learning is an off-policy RL algorithm used to find optimal action-selection policies in various optimization problems. It updates its estimations of future rewards based on the best possible action in the next state regardless of the policy currently followed. During the training, the agent stores the expected rewards in a Q-table for each state-action pair \((s, a)\). Through iterative interactions with the environment, the agent updates its Q-table and learns to choose the optimal actions to maximize the overall network performance.
At each step, the agent observes the current state \(s\)of the environment.  The agent selects action \(a\) based on this observation. The environment then moves to a new state \(s'\) as a consequence of the action chosen. Finally, the agent receives a reward signal \(r\) that reflects the outcome of the action in the new state. Q-learning algorithm aims to discover an optimal policy \(\pi^*\) that maximizes the long-term expected reward. To achieve this, the Q-function \(Q(s, a)\) is used to represent the total expected reward for taking an action \(a\) in state \(s\). The reward feedback enables the agent to refine its policy, leading to better decision-making.
Note that, the agent begins with no prior knowledge of the network, and  Q-values are initialized to zero. To balance exploration and exploitation, Q-learning employs an \(\varepsilon\)-greedy approach. The exploration factor, \(\varepsilon\), ranges from 0 to 1. In an unknown environment, \(\varepsilon\) is initially set high (\(\varepsilon \approx 1\)), encouraging the agent to explore various actions. As the agent learns, \(\varepsilon\) gradually decreases, shifting the focus towards exploiting actions with the highest known Q-values. To implement this, the agent generates a random value \(V\) between 0 and 1. If \(V < \varepsilon\) then the agent explores a new action, otherwise, the agent exploits its current knowledge by selecting the action with the highest Q-value. This iterative process of exploration and exploitation, guided by \(\varepsilon\), allows the agent to refine its understanding of the environment and ultimately converge towards the optimal policy. During the training, the Q-value is updated according to the Bellman equation expressed as

\begin{equation}
 Q^{\text{new}}(s, a) \leftarrow Q(s, a) + \alpha [r(s, a) + \gamma \max_{a'} Q(s', a') -Q(s, a)],
\end{equation}
\noindent where \(\alpha \in (0, 1]\) is the learning rate, and it controls how much the agent learns from a single experience when updating the Q-value. High \(\alpha\) leads to rapid learning and potentially inaccurate results, while low \(\alpha\) ensures accuracy at the cost of slow convergence. 
Moreover, \(\gamma \in (0, 1]\) is a discount rate ensuring convergence of the expected total reward. The process iterates until the Q-values converge, indicating that the optimal policy \(\pi^*\) is found or a maximum iteration limit is reached. In other words, once the learning is complete, the Q-learning algorithm outputs an optimal policy \( \pi^* \), which specifies the action \( a \) for each state \( s \) that maximizes the Q-value \( Q^*(s, a) \):
\begin{equation}
\pi^* = \max Q^*(s, a).
\end{equation}

\subsection{SARSA} \label{sec:Sarsa}


SARSA is an on-policy algorithm that updates Q-values based on actions taken by the agent within its current policy. The update process for the action-value function  \(Q(s,a)\) can be expressed as

\begin{equation}
Q(s,a) \leftarrow Q(s,a) + \alpha [r(s,a) + \gamma Q(s', a') - Q(s, a)].  
\end{equation}
In SARSA, the agent interacts with the environment by taking actions and observing the resulting state transitions, rewards, and updating its policy based on the next state and action estimates. At each state \(s\), the agent selects an action \(a\) based on an exploration policy, i.e., \(\varepsilon\)-greedy. The agent then receives a reward \(r\) and transitions to a new state \(s'\). The core of SARSA lies in its update rule, which refines the estimated Q-value for the state-action pair \((s, a)\). The updated process incorporates the immediate reward \(r(s, a)\), the estimated value of the next state \(Q(s', a')\), and a learning rate \(\alpha\) to balance immediate reward and long-term value. SARSA uses the actual next action \(a'\) taken by the policy during the update, as opposed to Q-learning which considers the maximally valued action in the next state.
\section{Results}\label{sec:Reslts}We consider the system model described in Section \ref{sec:Pro} {in an indoor environment of 5$m$ $\times$ 5$m$ $\times$ 3$m$}.  Four LED-based APs (\(L=4\)) are implemented on the ceiling with half-power semi-angel equal $60^0$. An IRS mirror array is mounted on one wall and contains 5 $\times$ 5 reflective elements. The mirror reflectivity is equal to 0.95 and each mirror has an effective area of 25 cm $\times$ 10 cm unless specified otherwise. On the communication plane, five active users (\(K=5\)) are randomly distributed on the communication floor. The receiver photodetector responsivity, optical filter gain, and the physical area are set to 0.4 A/W, 1 and 20 $mm^2$, respectively. In addition, the photodetector bandwidth is equal to 20 MHz. 
The RL agent is located in the CU connected to APs and the IRS mirror array, enabling the agent to make intelligent association decisions across the environment. The simulation is implemented in Python 3.8 and a customized Gym environment designed specifically for our IRS-aided VLC network. The assignment of the APs and mirrors is solved using Q-learning and SARSA algorithms, and the performance of these algorithms is evaluated.

\begin{figure}[t]
\vspace{-10pt} %
    \centering
    \includegraphics[width=0.5\textwidth]{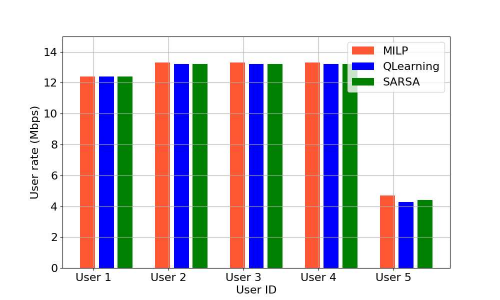}
\caption{Achievable user data rate using Q-Learning, SARSA and MILP. Transmitted optical power is 5 W.} 
\label{f33}
\end{figure}

Fig. \ref{f33} demonstrates the optimization problem solutions using the RL algorithms compared to the optimal solution of mixed integer linear programming (MILP), which is a model used as in \cite{elgamal2021reinforcement} to solve the formulated optimization problem.   The Q-learning and SARSA algorithms achieve near-optimal solutions without prior knowledge of the environment compared to the MILP, results that rely on full knowledge of the environment. Note that, the Q-learning algorithm allocates resources to each user based on the maximum aggregate data rate, i.e., rewards, while SARSA is an on-policy algorithm where it learns directly from the consequences of its actions, avoiding actions that might lead to negative consequences. The results also show that users experience different QoS as they differ in channel gain, i.e., their locations in relation to the APs and IRS.

Fig. \ref{fig4} presents the sum rate versus the transmitted optical power in various scenarios. The figure shows that the data rate increases as the transmitted optical power increases. However, the transmitted power is subject to illumination and eye safety regulations. Moreover, the proposed joint optimization of the IRS and AP associations using RL in our network improves the sum rate by 66\% compared to a scenario where no IRS is deployed and 45\% compared to a scheme where the IRS elements are assigned to the users based on the distance.

Fig. \ref{fig5} shows the achievable data rates of the proposed schemes versus the transmitted optical power in the presence of blocking objects relatively close to the users. The Matern hardcore point process (MHCP) model is used to simulate the position of static blockages within the room. In our model, a human is modeled as a cylindrical shape with a diameter of  0.3 m and a height of  1.65 m \cite{jacob2013fundamental}.
Naturally, the data rate decreases as the number of blockages increases. However, the use of the IRS and its association to the users ease the blockage problem in OWC. It is shown that deploying a single mirror array achieves data rates of up to 9 Mb/s under 2 blockages and 5.8 Mb/s under 3 blockages at a 2 W transmit power. Moreover, deploying two mirror arrays at the same transmit power achieves data rates of 10.2 Mbps under 2 blockages and 6.4 Mbps under 3 blockages, resulting in data rate increases of  10\% and 13\% compared to the single mirror array scenarios under 2 and 3 blockages, respectively. This enhancement is due to the increased signal reflection paths provided by the mirror elements.
{\setlength{\parskip}{0pt}
\begin{figure}[t]
    \vspace{-18pt} %
    \centering
    \includegraphics[width=0.5\textwidth]{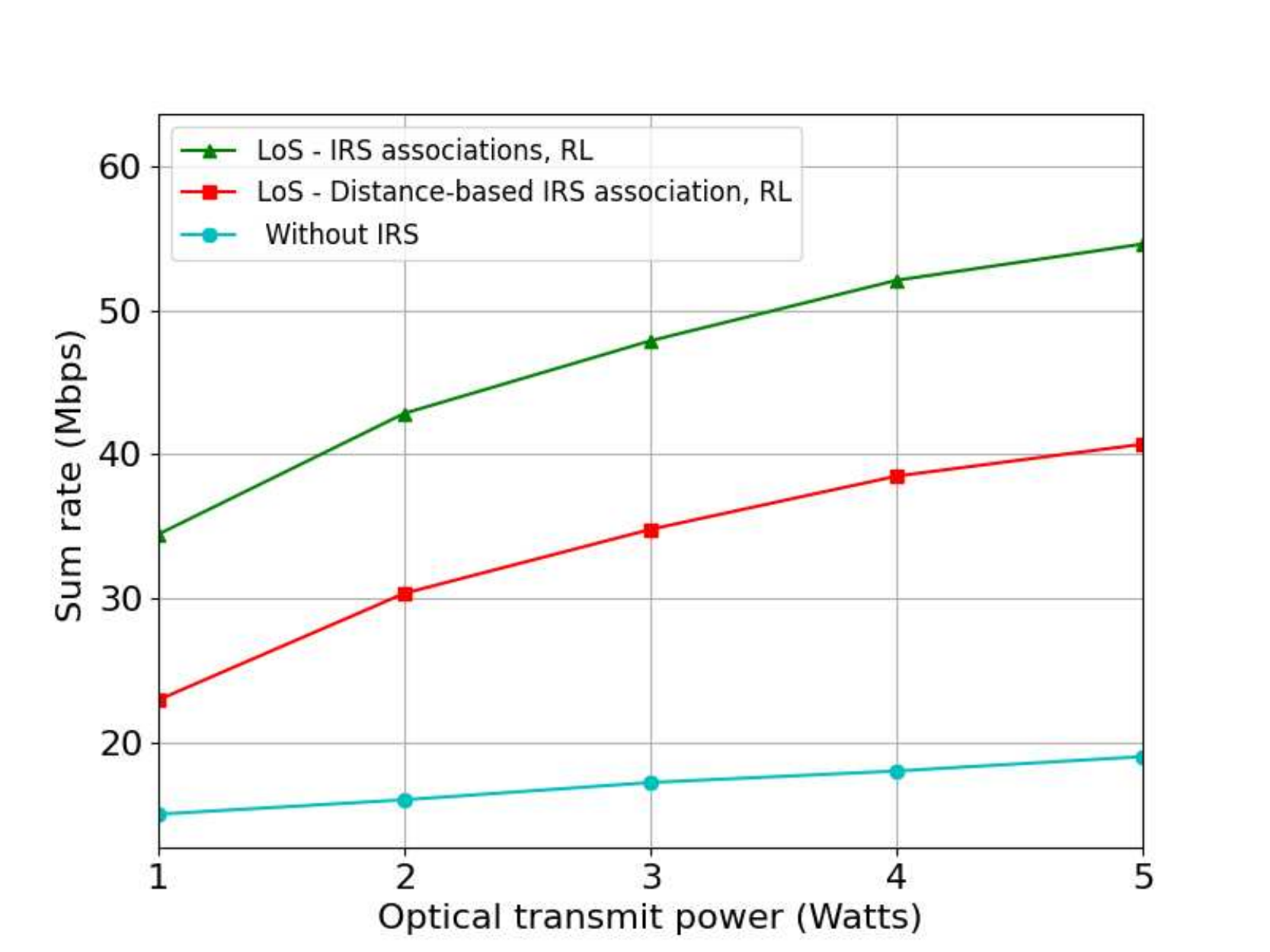}
     \captionsetup{justification=raggedright, singlelinecheck=false}
\caption{Sum rate versus the transmitted optical power.}
\vspace{-10pt}
\label{fig4}
\end{figure}
\begin{figure}[t]
 \vspace{-10pt}
    \centering
    \includegraphics[width=0.5\textwidth]{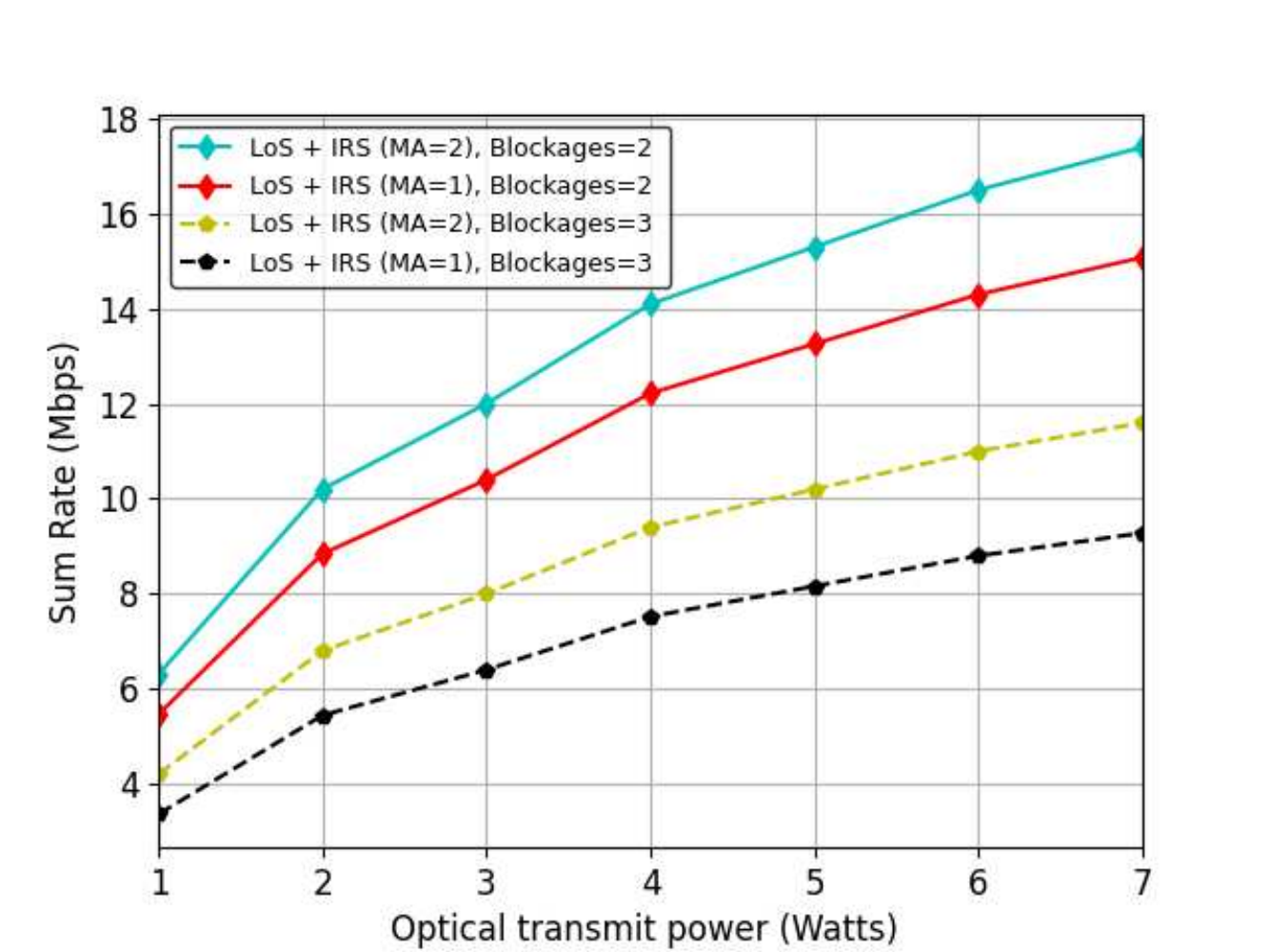}
\caption{ Sum rates of the proposed schemes versus the transmitted optical power.}
\label{fig5}
\end{figure}
\section{Conclusion}
This paper investigated the performance of IRS-aided OWC system composed of multiple LED-based APs serving multiple users and an array of mirrors acting as the IRS. First, an optimization problem was formulated to maximize the sum data rate by jointly optimizing the association of the APs and IRS mirrors to the users. The formulated optimization problem was solved using two RL algorithms, Q-learning and SARSA. The results demonstrated the optimality of the proposed RL algorithms, and the effectiveness of the proposed approach in improving the sum rate of the system compared to a traditional scheme that optimizes the allocation of APs with distance-based IRS association. The results also demonstrated that the use of the IRS significantly improves the sum rate of the network and mitigates LoS blockage.
\section*{Acknowledgment}
This work has been supported in part by the Engineering and Physical Sciences Research Council (EPSRC), in part by the INTERNET project under Grant EP/H040536/1, and in part by the STAR project under Grant EP/K016873/1 and in part by the TOWS project under Grant EP/S016570/1. All data are provided in full in the results section of this paper.
\bibliographystyle{ieeetr}
\bibliography{References}

\end{document}